\documentstyle[12pt]{article}
\input math_macros
\def\roughly#1{\raise.3ex\hbox{$#1$\kern-.75em\lower1ex\hbox{$\sim$}}}

\begin{document}
\begin{titlepage}
\begin{center}
November, 1992      \hfill       LBL-33122 \\
                    \hfill       UCB-PTH-92/38 \\
                    \hfill       hep-ph/9211233 \\

\vskip .2 in
{\large \bf  Anomalous prompt photon production in hadronic
collisions at low-$x_T$}
\footnote{This work was supported by the Director, Office of Energy
Research, Office of High Energy and Nuclear Physics, Division of High
Energy Physics of the U.S. Department of Energy under Contract
DE-AC03-76SF00098.}
\vskip .3 in
       {
      {\bf Damien Pierce} \\
           \vskip 0.5 cm

      {\em Department of Physics \\
          University of California, Berkeley \\
          and\\
          Theoretical Physics Group \\
          Lawrence Berkeley Laboratory \\
          1 Cyclotron Road, Berkeley, CA 94720} }
\end{center}
\vskip 0.3 in
\begin{abstract}
We investigate the discrepancy that exists at low-$x_T=2p_T/\sqrt{s}$
between the next--to--leading order QCD
calculations of prompt photon production and the measured cross
section. The central values of the
measured cross section are of order 100\% larger than
QCD predictions in this
region. It has been suggested that the bremsstrahlung contribution
may account for this discrepancy.
The quark fragmentation function $D_{\gamma/q}(z)$ has not been
measured and an exactly known asymptotic form is normally used in
calculations. We examine the
effect of much larger fragmentation functions on the QCD
predictions. After illustrating the effect of the large fragmentation
functions in some detail for recent CDF data at $\sqrt{s}$=1.8~TeV,
we perform a $\chi^2$ fit to 8 prompt photon data sets ranging in
CMS energy from 24~GeV to 1.8~TeV.  While a large fragmentation
function normalization may
prove to play an important role in resolving the discrepancy,
the present theoretical and experimental
uncertainties prevent any definite normalization value
from being determined.
\end{abstract}
\end{titlepage}

\renewcommand{\thepage}{\roman{page}}
\setcounter{page}{2}

\vskip 1in

\begin{center}
{\bf Disclaimer}
\end{center}

\vskip .4 in

\begin{scriptsize}
\begin{quotation}
This document was prepared as an account of work sponsored by the United
States Government.  Neither the United States Government nor any agency
thereof, nor The Regents of the University of California, nor any of their
employees, makes any warranty, express or implied, or assumes any legal
liability or responsibility for the accuracy, completeness, or usefulness
of any information, apparatus, product, or process disclosed, or represents
that its use would not infringe privately owned rights.  Reference herein
to any specific commercial products process, or service by its trade name,
trademark, manufacturer, or otherwise, does not necessarily constitute or
imply its endorsement, recommendation, or favoring by the United States
Government or any agency thereof, or The Regents of the University of
California.  The views and opinions of authors expressed herein do not
necessarily state or reflect those of the United States Government or any
agency thereof of The Regents of the University of California and shall
not be used for advertising or product endorsement purposes.
\end{quotation}
\end{scriptsize}

\vskip 2in

\begin{center}
\begin{small}
{\it Lawrence Berkeley Laboratory is an equal opportunity employer.}
\end{small}
\end{center}

\newpage
\renewcommand{\thepage}{\arabic{page}}
\setcounter{page}{1}

\section{Introduction}
Prompt photon production in hadronic collisions is an important
testing ground for QCD.
Recently, measurements of inclusive prompt photon production at
CMS energy $\sqrt{s}=$1.8~TeV have been published by the CDF
collaboration, and a comparison with QCD
calculations at next--to--leading order has been carried out
\cite{CDF}. While the QCD predictions give good qualitative
agreement, the central values of the measured
cross section in the low-$p_T$ region are of order 100\% larger
than standard QCD calculations. The same discrepancy exists between
theory and experiment for data taken by UA1 and UA2 at the CMS
energy $\sqrt{s}$=630~GeV.

The NLO QCD calculation consists of three parts.
$$\sigma = \sigma^{LL} + \sigma^{NLL} + \sigma^{ANOM}$$
where the three terms indicated are the leading logarithm
(improved Born approximation) contribution, the next-to-leading
logarithm piece, and the so-called ``anomalous'' part,
in which a photon is emitted collinearly
from an outgoing parton. This is just the bremsstrahlung process.
While the first two parts of the QCD calculation are reliably
calculated
(given the fact that the quark and gluon distribution functions are
reasonably well constrained)
the third, the bremsstrahlung
process, is not. The calculation of the bremsstrahlung process
involves an unmeasured set of non-perturbative functions, the
parton $\rightarrow$ photon fragmentation functions. While there is
a standard parametrization \cite{Owens}
for these functions used by most theorists in QCD calculations, this
parametrization is only valid at asymptotically large energy scales
where the effects of boundary conditions are vanishing.
In this paper we examine the effect of large fragmentation
functions on the QCD predictions of inclusive prompt photon
production and compare with experimental data.

This paper is organized as follows. In Section 2 we discuss the
NLO QCD calculation of
the bremsstrahlung component of prompt photon production and we
elaborate on certain
aspects of the fragmentation functions involved in this calculation.
In Section 3 we first illustrate the existing discrepancies between
theory and experiment for the UA1 and UA2 data at $\sqrt{s}$=630~GeV
and CDF data at $\sqrt{s}$=1.8~TeV.
Concentrating on the CDF data, we examine the effects of
the theoretical uncertainties due to the choice of scales and
distribution functions on the discrepancy. We then go on to
show the results of the QCD prompt photon calculation including
larger fragmentation functions.
We compare our results with the standard calculations, and with the
CDF data.
We again examine the effect of changing the scales and parton
densities on the theoretical predictions. In Section 4 we proceed
to compare the results of the calculation with and without a large
set of fragmentation functions to data at center of mass energies
$\sqrt{s}$=1800, 630, 546, 63, and 24 GeV. A $\chi^2$ analysis is
performed and the results are discussed.
Section 5 is reserved for the conclusions.

\section{Bremsstrahlung calculation; Fragmentation functions}
The anomalous part of the QCD calculation of the inclusive prompt
photon production differential cross section
$\sigma(AB\rightarrow \gamma + X)$
is given at the leading logarithm level by
\begin{equation}
\sigma^{ANOM} = \sum_{abcd}\int F_{a/A}(x_a,\mu_F^2)
\,F_{b/B}(x_b,\mu_F^2)\,D_{\gamma/c}(z,M_f^2)\;{d\hat{\sigma}\over dv}
\,dx_a\,dx_b\,dz\,dv.
\end{equation}
In this equation $v=1+{\hat t}/{\hat s}$, where $\hat t$ and
$\hat s$ are the Mandlestam variables of the subprocess,
and $x_a,\ x_b,$ and $z$ are the fractional
momenta of the initial state partons and the photon, respectively.
The subprocess cross section $d\hat{\sigma}/dv$,
which describes the parton level scattering process
$ab\rightarrow cd$, is convoluted with the parton distribution
functions $F_{a/A}$ and $F_{b/B}$ and the
parton~$\rightarrow$~photon fragmentation functions $D_{\gamma/c}$.
The three scales in the calculation are the renormalization
scale $\mu_R$, associated with the strong coupling constant,
the factorization scale $\mu_F$ where the parton distribution
functions are evaluated, and the fragmentation scale, $M_f$,
associated with the parton~$\rightarrow$~photon fragmentation process.
At sufficiently high order in perturbation theory,
the total calculated cross section should be insensitive to the
chosen values of these scales.

The fragmentation and distribution functions are nonperturbative
objects and are treated as measured inputs within the framework of
perturbative
QCD\footnote{Given a set of functions at some scale the evolution
of them to other energy scales is perturbatively calculable.}.
While many experiments such as deep inelastic scattering have
given rise to a fairly well constrained set of quark and gluon
distribution functions, the parton~$\rightarrow$~photon
fragmentation functions have not been
measured. However, owing to the pointlike nature of the photon
these fragmentation functions differ essentially from the
parton distribution functions in that at asymptotically large scales
the fragmentation functions are perturbatively calculable.
The asymptotic form for the quark and gluon fragmentation functions,
which includes the summation to all orders in perturbation theory
of soft and collinear gluon emission at leading logarithm level, is
parametrized as follows \cite{Owens}
\begin{equation}
zD^{LL}_{\gamma/q}(z,Q^2)=F\left[
{e_q^2(2.21-1.28z+1.29z^2)z^{0.049}\over1-1.63\ln(1-z)}
+0.0020(1-z)^{2.0}z^{-1.54}\right]
\end{equation}
\begin{equation}
zD^{LL}_{\gamma/g}(z,Q^2)=F{0.194\over8}(1-z)^{1.03}z^{-0.97},
\end{equation}
where
$F=(\alpha/2\pi)\ln\left(Q^2/\Lambda^2_{
\mbox{$\scriptscriptstyle QCD$}}\right)$.
Theorists generally use this asymptotic form when doing QCD
calculations
of prompt photon production and comparing to the experimental
data \cite{Owens,Aurenche,Baer}.
However, the boundary conditions on the evolution equations may be
important at scales where experimental data has been taken.
The boundary conditions on the evolution equations are simply the
initial fragmentation functions at some scale $Q_0$, where $Q_0$ is
defined to be the scale above which the perturbative evolution
equations are assumed
to be valid. We take this scale to be 2~GeV.
We consider the possibility
that the physics at scales below $Q_0$ builds up a set of functions at
$Q_0$ with a larger normalization than that of the asymptotic form
evaluated at the starting scale.

In order to make the analysis tractable we consider starting quark
and gluon fragmentation functions
which have the same shape as the asymptotic forms
but which have normalizations $N_0$ times that of the asymptotic forms
evaluated at the scale $Q_0$.
Of course, as the scale increases these large fragmentation
functions evolve
and approach the asymptotic form normalization. In figure 1a we show
the up-quark fragmentation function both for the asymptotic form
and for a large fragmentation function with $N_0=20$. The
asymptotic form grows logarithmically with the scale and this
factor has been divided out in fig.1a.
We show in this plot the large fragmentation function evaluated
at the starting scale 2~GeV, the scale 10~GeV and the scale 100~GeV.
In figure 1b we show the normalization of the large fragmentation
function
vs. the scale $Q$, for values of $N_0$ equal to 10, 20, and 30.
Because of isolation cuts, many experiments are only
sensitive to large values of fragmentation function fractional
momenta. For this reason, the relative normalization $N(Q^2)$ shown
in fig.1b is given by
\begin{equation}
N_0(Q^2)={\int_{\scriptscriptstyle 0.85}^{\scriptscriptstyle 1}\,
D^{LL}_{N_0}(z,Q^2)\,\,dz \over
\int_{\scriptscriptstyle 0.85}^{\scriptscriptstyle 1}\,
D^{LL}_{\rm asymp}(z,Q^2)\,\,dz}
\end{equation}
where the denominator is the asymptotic form fragmentation function
of eq.2. Note that the relative normalization
decreases rapidly in the first 10 or 20~GeV beyond the starting scale
$Q_0$. This feature is well suited for resolving the discrepancy
between the QCD prompt photon predictions and experimental data at
low transverse momenta. If the starting fragmentation function
normalization is much larger than the asymptotic form normalization,
the smaller $p_T$ values will probe this large normalization, while
at larger $p_T$ values most of the initial large normalization will
have evolved away. Thus, in the low-$p_T$ region the predicted
cross section will markedly increase relative to the standard
calculations, while at higher photon transverse momenta
we can expect results similar to the standard QCD predictions.

It is of interest to estimate the largest initial fragmentation
function normalization that may be
physically relevant. An order of magnitude estimate for an upper
bound comes
from demanding that at the starting scale $Q_0$ the total
number of photons fragmented from some parent quark
with fractional momenta above some cut-off $z_0$ be less than some
small number of ${\cal O}(\alpha/\alpha_s)$, say $1\over40$,
times the total number of
hadrons that are fragmented by the quark. We show in figure 2 the
maximum normalization $N_{max}$ as a function of $z_0$ coming from
this constraint.
We assume a hadron multiplicity of $\sqrt{2E}$ for a parent quark of
CMS energy
$E$, in~GeV units, and have fixed the parent quark energy at 3~GeV.
Taking $z_0$ to be 0.3, we see from fig. 2 that we can tolerate an
initial normalization which is $\sim$20 times larger than that of
the asymptotic form.

Another more quantitative bound can be estimated by examining the
parton~$\rightarrow$~photon process in the crossed channel, namely the
photon~$\rightarrow$~quark
fragmentation function. In this case the photon structure function
in the non-perturbative regime can estimated by appealing to vector
meson dominance (VMD) arguments. In ref.\cite{Pilon} the
$\gamma\rightarrow$~vector meson
fragmentation function is hypothesized to be equal to the
$\gamma\rightarrow \pi^0$
fragmentation function, which is extracted from fits to
$e^+e^-\rightarrow\pi^0X$ data.
We find from the analysis of ref.\cite{Pilon} that using the VMD
hypothesis to specify the fragmentation function boundary
conditions limits $N_0$ to
be less than 2 at leading order. However, due to the many theoretical
uncertainties involved in the VMD procedure, we will not respect
this limit. Rather, we will go ahead and vary $N_0$ freely in order
to determine whether
the present experimental prompt photon data can constrain it directly.

We will find that in order to account for a substantial increase in
the prompt photon production cross section at low transverse
momenta we will need to consider initial fragmentation function
normalizations which are in some cases 30 or more times larger than
the normalization of the asymptotic form (which is probably
unreasonably large). At first sight this may be surprising since at
large center of mass energies and low-$p_T$ it is well known that
the bremsstrahlung contribution can dominate the other photon
production processes. However, in order to reduce the background
photons from $\pi^0$ and $\eta$ decay the data samples are taken
with the imposition of an isolation cut. These isolation cuts
effectively remove most of the bremsstrahlung events
from the data sample. Typically only photon fractional momenta
values of $z \roughly{>}0.85$ are allowed. Additionally, at the lowest
$p_T$ values measured in the experiments the fragmentation function
normalizations will have evolved considerably down from any large
starting normalization.

\section{Results}
We use the algorithm of Baer {\it et al.} \cite{Baer} to calculate the
next-to-leading order QCD cross section for prompt photon production
in hadronic collisions. See their paper and references therein for the
relevant formulae. The calculation involves a Monte Carlo integration
wherein various isolation criteria are easily implemented.
To simplify our analysis we choose to set the renormalization
scale equal to the factorization scale, $\mu^2=\mu_R^2=\mu_F^2$.

In figure 3 we compare our results for prompt photon production
with CDF data \cite{CDF} taken at $\sqrt{s}$=1.8~TeV, and with UA1
and UA2 data  \cite{UA1,UA2_92} taken at
630~GeV. Here we are using the standard asymptotic form
parametrization of the quark and gluon fragmentation functions
given in eqs.(2)
and (3) ({\it i.e.} $N_0$=1). We show the results for the scales
$\mu^2=M_f^2=p_T^2$/4 and the distribution functions
Morfin \& Tung Set 1 (M-T1) \cite{MorfTung}.
Both the CDF data and the corresponding calculation include an
isolation cut, such that the total hadronic energy inside a cone of
size $\Delta R =\sqrt{\Delta \eta^2 + \Delta \phi^2}$=0.7
centered on the photon be less than 2~GeV ($\eta$ and $\phi$
are the psuedo-rapidity and azimuthal angle, respectively).
The UA1 data and the corresponding
calculation include an isolation cut such that the total hadronic
energy inside a cone around the photon of size $\Delta R$=0.4 be
less than 10\% of the photonic energy. The isolation criteria for
the UA2 data shown in fig. 3 is not as simple, but in an earlier
UA2 publication of prompt photon data \cite{UA2_88} it was
demonstrated that
the data sample was consistent with having assumed an isolation
cut such that the hadronic energy in a cone around the photon of size
$\Delta R$=0.78 be less than 3.5~GeV. We use this isolation criteria
in our calculation. The UA1 and UA2 isolation criteria give very
similar results and we show the curve corresponding to the UA1
isolation criteria in fig. 3. The data samples have the overall scale
normalization uncertainties shown in the figures. The error bars in
the figures include statistical and systematic errors added in
quadrature. The discrepancy in the low-$p_T$ region is apparent for
all three data sets.

The discrepancy is more easily seen on a linear scale. In
figure 4a we show the relative difference between data and theory,
(data-theory)/theory, versus $p_T$ for the CDF data. This figure
illustrates the theoretical uncertainty due to changing the scales
and distribution
functions. The reference theory calculation on this
plot includes M-T1 distribution functions and scales equal
to $p_T^2$. The results using the DFLM260 \cite{DFLM} distribution
functions are shown and give about a~20\% larger cross section at
low-$p_T$,
decreasing to less than 10\% of the M-T1 result at large $p_T$.
Also in this figure we see that changing the scales to
$\mu^2=M_f^2=p_T^2/4$ increases the results of the predictions
by 10-15\% (for a given set of distribution functions).
The results are relatively insensitive to changes in the factorization
scale $M_f^2$. For standard initial normalizations ($N_0=1$) the
results change by $\roughly{<}$~5\% when considering $M_f^2=p_T^2$ and
$M_f^2=p_T^2/4$ or $M_f^2=4*p_T^2$. At larger initial fragmentation
function normalization values the fragmentation scale dependence is
larger, but generally it is $\roughly{<}$~10\% for starting
normalization values less than 40.

The discrepancy is readily observed in fig. 4a for the case of
$p_T^2$ scales and M-T1 distribution functions. For these choices
the central values of the three lowest $p_T$ bins are 60, 51 and
117\% larger than the theoretical calculation, and have a
$\chi^2$/d.o.f. of 5.3. For comparison, the central values of the
next lowest three $p_T$ bins average less than 9\% larger than the
theoretical curve. The choices corresponding to the
largest theoretical curve are $p_T^2/4$ scales and DFLM
distribution functions. In this case there is certainly no glaring
discrepancy in the low-$p_T$ region, although the central value
of the lowest $p_T$ bin is still 80\% larger than the prediction.

In figure 4b we illustrate the discrepancy between data and theory
for the UA1 and UA2 data at $\sqrt{s}$=630~GeV. As in fig.4a the
largest low-$p_T$ region discrepancy
corresponds to the theoretical calculation which includes M-T1
parton densities and $p_T^2$ scales. For these choices the lowest
four $p_T$ bins average 97\% larger than the expectation. The
discrepancy is least, as for the CDF data,
in the case of DFLM parton densities and scales equal to $p_T^2/4$.
In this case the $\chi^2$/d.o.f. of the lowest four $p_T$ bins is 3.2.

We show in figure 5a the qualitative improvement in the comparison of
CDF data and theory if we include a large initial normalization for
the fragmentation functions. We show the curves for $N_0$=30 and 40
for the case of M-T1 distribution functions and $p_T^2$ scales.
The large $N_0$ curves qualitatively account for the rise in the
cross section in the low-$p_T$ region while retaining the good
description at large $p_T$. The best fit (defined by the smallest
$\chi^2$; see Sec.4) for the initial normalization $N_0$ for this
choice of scales and distribution functions is $N_0$=34.
Of course, the best fit value for the starting normalization
$N_0$ will change with the distribution function and scale choices.
If the scale $\mu^2$ is chosen to be smaller ({\it e.g.}
$\mu^2=p_T^2/4$) the QCD prediction will increase overall, thus a
smaller normalization will be favored. As an illustration, we show
in figure 5b the standard ($N_0$=1) curve and the best fit
($N_0=17$) curve for the scale choice $\mu^2=M_f^2=p_T^2/4$.
Similarly, as the DFLM distribution functions give larger cross
sections, they will
also give smaller values of the best fit starting normalization.
Figure 5c shows the DFLM curve for the standard calculation
($N_0=1$) and the best fit initial normalization in this case,
$N_0=17$, for the choice of scales $\mu^2=M_f^2=p_T^2$.

If we move from the choice of distribution
function and scale of fig. 5a to the one of DFLM distribution
functions and
the scale choice $\mu^2=p_T^2/4$, we can expect an even smaller best
fit initial fragmentation function normalization value, as both of
these choices increase the theoretical values of the cross section
relative to the choices of fig. 5a. We find the best fit value in
this case is $N_0=4$. However, it is perhaps worth noting that this
case gives the worst fit to the data. The central values of all the
data points (except for the three lowest $p_T$ bins) fall below the
theoretical calculation (even in the case $N_0=1$) and
correspondingly the $\chi^2$ value for this choice of scales
and distribution functions is the largest $\chi^2$ of the four
choices considered.

We have seen that while varying the choice of scales and
distribution functions in the theoretical calculation at
$\sqrt{s}$=1.8~TeV
the best fit value for the initial normalization $N_0$ varies from
4 to 34. The uncertainty in the QCD calculation is too great to
yield a well constrained value for the normalization. However, even
if the QCD calculation were under better control, we will see in
the next section that the experimental errors are too large to lend
a best fit normalization value any statistical significance.

\section{$\chi^2$ analysis}
In order to understand the statistical significance of the qualitative
improvement achieved by stipulating a large initial fragmentation
function normalization, and to see if such a large fragmentation
function is consistent with other lower energy experimental data,
we vary the initial fragmentation function normalization and perform
a $\chi^2$ analysis. In most of the data sets there is an overall
scale normalization uncertainty. In such cases we evaluate the
$\chi^2$/d.o.f. by varying the overall normalization of all the
data points and
finding the normalization $N$ which renders the $\chi^2$ formula
\begin{equation}
\chi^2{\rm /d.o.f.} = {1\over L}\sum_{i=1}^L
{\left(\sigma^{exp}(p_{T_i})-\sigma^{theory}(p_{T_i})\right)^2\over
\left(\Delta \sigma^{exp}(p_{T_i})\right)^2} + {(N-1)^2\over
\Delta N^2}
\end{equation}
a minimum. $\Delta N$ is the experimental normalization error.
For $\Delta \sigma^{exp}$ we add the statistical and systematic
errors in
quadrature. If part of the systematic error is correlated
the true $\chi^2$ values for that data sample will be larger than
the ones evaluated using eq.(5).
In figures 6 we show the $\chi^2$ distributions
for three representative data sets;
the CDF data at $\sqrt{s}=$1.8~TeV, the combined $\chi^2$ for the
UA1 and UA2 data sets at $\sqrt{s}=$630~GeV, and the NA24 data taken
at $\sqrt{s}=$23.76~GeV. In fig.6a the scales are set to
$\mu^2=M_f^2=p_T^2$ and the DFLM distribution functions are used.
These three data sets show a favored initial fragmentation function
normalization value much larger than the standard value of 1 (the
values are
$N_0$=17,\ 24, and 19 respectively). However, for both the 1.8~TeV and
the 630~GeV data, the difference in the $\chi^2$ value at the
minimum and that for an initial normalization of 1 (or zero) is
less than 0.5. Thus, while the experimental data favors a large
initial fragmentation normalization, the
experimental errors are such that a standard initial normalization
(or even zero initial normalization) gives almost as good a fit.

For the low energy data set shown, the situation is quite different.
The reason the $\chi^2$ distribution corresponding to the
NA24 data set shows a strong dependence on the initial fragmentation
function normalization is that this data set has no isolation
criteria applied. The $\chi^2$ plot in fig.6a shows the value
$N_0=1$ is actually ruled out at the 68\% confidence level. However,
by changing the scales and/or parton distribution functions in the
calculation the value $N_0$=1 can be accommodated. This is shown in
fig.6b, in which the scales in the calculation have been set to
$p_T^2/4$. All three distributions in fig.6b show minima
at smaller values of $N_0$ than in fig.6a, and in particular the NA24
data in this case shows a minimum at $N_0$=5. The $\chi^2$/d.o.f.
value corresponding to $N_0=1$ differs by less than 1 from the
$\chi^2$/d.o.f. value at the minimum.

We now discuss the combined $\chi^2$ plots for all of the data samples
considered. The data samples included are the CDF data at
$\sqrt{s}=$1.8~TeV and $\eta=0$ \cite{CDF};
the UA1 data at $\sqrt{s}=$630~GeV and $\sqrt{s}=$546~GeV, both for
$\eta=0$ and 1.1 \cite{UA1};
the UA2 data at $\sqrt{s}=$630~GeV, $\eta=0$ \cite{UA2_92};
R806 data at $\sqrt{s}=$63~GeV and $\eta=0$ \cite{R806};
and NA24 data at $\sqrt{s}=$23.76~GeV with $-.65<\eta<.52$
\cite{NA24}. Eight QCD calculations were performed for each of the
data sets corresponding to the
distribution function choices M-T1 and DFLM, and the scale
choices $\mu^2=n*p_T^2;\,M_f^2=m*p_T^2$ for
$(n,m)$=(1,1), (1,4), $({1\over4},{1\over4})$, and $({1\over4},1)$.
We show in figure 7a the $\chi^2$ distribution as a function of the
initial fragmentation function normalization $N_0$ for both sets
of structure functions considered and for the scale choice
$\mu^2=M_f^2=p_T^2$.
In this figure the initial normalization values corresponding to the
lowest $\chi^2$ values are $N_0$=24 and $N_0$=19 for the
M-T1 and DFLM distribution function choices, respectively.
In figure 7b we show the results for the scales equal to $p_T^2/4$.
Here the minimum $\chi^2$ values correspond to the normalizations
$N_0=8$ and $N_0=4$ for the M-T1 and DFLM distribution
function choices, respectively. The $\chi^2$ distributions
corresponding to the same $\mu^2$ scale choices as in figs.7 but
different $M_f^2$ scale choices give results similar to those shown
in figs. 7.

The $\chi^2$ plots all show very mild dependence on the initial
normalization $N_0$. A mild dependence for the higher energy data
sets can be explained as follows. In the case of the CDF data the
theoretical predictions and the experimental data
agree within the (large) experimental errors for 11 out of 12 of the
highest $p_T$ data points. The data set, even with a discrepancy
for the two or three lowest $p_T$ data points, has a low
$\chi^2$/d.o.f. The $\chi^2$/d.o.f. value changes very little
if the theory is modified in such a way that the three lowest $p_T$
data points are also brought into agreement.

In none of the eight cases considered is the difference between the
$\chi^2$ values corresponding to the standard normalization $N_0=$1
and the normalization at the minimum of the $\chi^2$ plot greater
than 0.4. Thus, because the experimental errors are large, the
quantitative improvement in the comparison of theory and data
obtained by including a larger
fragmentation function normalization is small. However, we do find it
somewhat suggestive that in the individual $\chi^2$ distributions
for all of the data sets considered and for all choices of scales
and parton densities considered (64 plots) the minima were more
often than not at initial fragmentation normalization values greater
than 16.

\section{Conclusions}
We find that including large boundary conditions in the evolution
equations of the {\it quark $\rightarrow$ photon} and {\it gluon
$\rightarrow$ photon} fragmentation
functions may play an important role in resolving the discrepancy
between next--to--leading order QCD calculations and experimental data
of the inclusive prompt photon production cross section
in the low-$x_T$ region.
The fragmentation function evolutions were performed at leading
order here. Perhaps including large initial fragmentation function
normalizations in a full next--to--leading analysis would lead to
a better constrained set of best fit
initial normalization values.
The inclusive prompt photon data sets considered here favor a larger
initial normalization, but the theoretical
uncertainties arising from scale variations and parton density
uncertainties are too large to give a definite value.
A $\chi^2$ analysis reveals that, additionally, the experimental
errors are too large to give any statistical
significance to what appears to be a qualitative improvement in the
comparison of theory and data by including a larger fragmentation
function normalization. In order to satisfactorily resolve the
discrepancy, both theoreticians and experimentalists will have to
further refine their respective results. On the theoretical side,
more tightly constrained proton structure functions and higher
order calculations would certainly stabilize the predictions.
On the experimental side we note that if $N_0$ is large
the photon production cross section is particularly sensitive to
the isolation criteria (either the isolation cone size or the
amount of hadronic energy allowed in the cone). A constraint on the
initial fragmentation normalization could come from an experimental
determination of the dependence of the photon production cross
section on the isolation criteria.
\vskip 24pt
\hbox{{\large {\bf Acknowledgements}}}
\vskip 12pt
We would like to thank Dr. Ian Hinchliffe for suggesting
this project to us, and for many useful discussions. In addition
we would like to acknowledge the help of Marjorie Shapiro,
Les Nakae and John Huth of the CDF collaboration.
This work was supported by the Director, Office of Energy Research,
Office of High Energy and Nuclear Physics, Division of High Energy
Physics of the U.S. Department of Energy under contract
DE-AC03-76SF00098 and in part by a University of California
Department of Education fellowship.

\newpage
\hbox{\large \bf Figure captions}
\vskip 16pt

\hbox{\large \bf Figure 1}

(a) The up quark fragmentation function versus $z$. The curves
are shown for $N_0$=20 at scales $Q_0=2$~GeV, and $Q$=10 and 100~GeV.
The asymptotic form ($N_0$=1) is also shown.

(b) The normalization of the up quark fragmentation function
versus the scale $Q$ for initial normalization values $N_0=$10, 20
and 30. The normalization is $N_0$ times the
asymptotic form normalization at the scale $Q_0=2$~GeV.
The dotted lines show the asymptotes at $Q_0$=2~GeV and $N$=1.

\vskip 16pt

\hbox{\large \bf Figure 2}
\noindent
The maximum initial normalization for the up quark fragmentation
function $N_{max}$ versus the photon fractional momenta cut-off
$z_0$, assuming the number of photons fragmented from a 3~GeV quark
with fractional momenta $z>z_0$ is ${1\over40}$ times the number of
hadrons fragmented by the quark.

\vskip 16pt

\hbox{\large \bf Figure 3}
\noindent
The inclusive prompt photon differential cross section
$E\,d^3\sigma/dp^3$ versus the photon transverse momenta. The data
is from CDF at $\sqrt{s}$=1.8~TeV and UA1 and UA2 at
$\sqrt{s}=$630~GeV. The curves are the QCD predictions
using Morfin-Tung Set 1 distribution functions and scales equal
to $p_T^2/4$. Note the experimental normalization uncertainties.

\vskip 16pt

\hbox{\large \bf Figure 4}
\noindent
The relative difference between data and theory,
(data-theory)/theory, is shown  vs. transverse momenta. The
reference theory calculation includes $p_T^2$ scales with M-T1 parton
densities. Also shown are the QCD results for scales $p_T^2/4$
and DFLM distribution functions.

(a) CDF data at $\sqrt{s}$=1.8 TeV.

(b) UA1 and UA2 data, $\sqrt{s}$=630 GeV.

\pagebreak

\hbox{\large \bf Figure 5}
\noindent
The relative difference between data and theory,
(data-theory)/theory, vs. transverse momenta for the CDF data at
$\sqrt{s}$=1.8~TeV. The reference theory calculations include the
standard asymptotic form fragmentation functions.

(a) M-T1 distribution functions and scales $\mu^2=M_f^2=p_T^2$
are used. Initial fragmentation function normalizations $N_0$=30 and
$N_0$=40 are shown. The best fit initial normalization value in this
case is $N_0$=34.

(b) The results for M-T1 distribution functions with scales
$\mu^2=M_f^2=p_T^2/4$. The best fit initial fragmentation function
normalization value $N_0$=17 is shown.

(c) DFLM parton densities are used, and the scales are set equal to
$p_T^2$. The best fit normalization value $N_0$=17 is shown.
\vskip 16pt

\hbox{\large \bf Figure 6}
\noindent
The $\chi^2$ distributions vs. the initial fragmentation
function normalization $N_0$ for three data sets; the CDF data
at $\sqrt{s}=1.8$~TeV, the combined $\chi^2$ for the UA1 and UA2 data
at $\sqrt{s}=$630~GeV, and NA24 data at $\sqrt{s}=23.76$~GeV. All
the data sets are at zero rapidity. The DFLM distribution functions
were used.

(a) The scales are set to $\mu^2=M_f^2=p_T^2$.

(b) The scales are set to $\mu^2=M_f^2=p_T^2/4$.

\vskip 16pt

\hbox{\large \bf Figure 7}
\noindent
The combined $\chi^2$ distributions vs. the initial fragmentation
function normalization $N_0$ for eight data sets; see text.

(a) The scales are set to $\mu^2=M_f^2=p_T^2$.

(b) The scales are set to $\mu^2=M_f^2=p_T^2/4$.

\end{document}